\documentclass{article}
\usepackage[misc]{ifsym}
\usepackage{hyperref}
\usepackage{graphicx}
\usepackage{amsmath}
\usepackage{booktabs}
\usepackage{multicol}
\usepackage{multirow}
\usepackage{mathptmx}
\usepackage{subcaption}
\usepackage{authblk}

\title{\Large\textbf{CryoMAE: Few-Shot Cryo-EM Particle Picking with Masked Autoencoders}}

\author{\normalsize\textbf{Chentianye Xu}\textsuperscript{1,†}, \textbf{Xueying Zhan}\textsuperscript{2,†}, \textbf{Min Xu}\textsuperscript{2,\Letter} \\ \textsuperscript{1} Language Technologies Institute, Carnegie Mellon University, Pittsburgh, PA 15213, USA \\ \textsuperscript{2} Computational Biology Department, Carnegie Mellon University, Pittsburgh, PA 15213, USA \\ \texttt{chentiax@cs.cmu.edu} \\ \texttt{xueyingz@andrew.cmu.edu} \\ \texttt{mxu1@cs.cmu.edu}}

\usepackage{geometry}
\date{\vspace{-0.8cm}}
\begin{document}
\maketitle

\begin{abstract}
Cryo-electron microscopy (cryo-EM) emerges as a pivotal technology for determining the architecture of cells, viruses, and protein assemblies at near-atomic resolution. Traditional particle picking, a key step in cryo-EM, struggles with manual effort and automated methods' sensitivity to low signal-to-noise ratio (SNR) and varied particle orientations. Furthermore, existing neural network (NN)-based approaches often require extensive labeled datasets, limiting their practicality. To overcome these obstacles, we introduce \textbf{cryoMAE}, a novel approach based on few-shot learning that harnesses the capabilities of Masked Autoencoders (MAE) to enable efficient selection of single particles in cryo-EM images. Contrary to conventional NN-based techniques, cryoMAE requires only a minimal set of positive particle images for training yet demonstrates high performance in particle detection. Furthermore, the implementation of a self-cross similarity loss ensures distinct features for particle and background regions, thereby enhancing the discrimination capability of cryoMAE. Experiments on large-scale cryo-EM datasets show that cryoMAE outperforms existing state-of-the-art (SOTA) methods, improving 3D reconstruction resolution by up to $22.4\%$.
    \\ \\ \noindent \textbf{Keywords:} Cryo-electron microscopy; particle picking; few-shot learning; masked autoencoder.
\end{abstract}

\footnotetext{\textsuperscript{†}Chentianye Xu and Xueying Zhan contributed equally to this work.}

\section{Introduction}
\label{sec:intro}

\begin{figure}[!htb]
  \centering
  \includegraphics[width=0.9\textwidth]{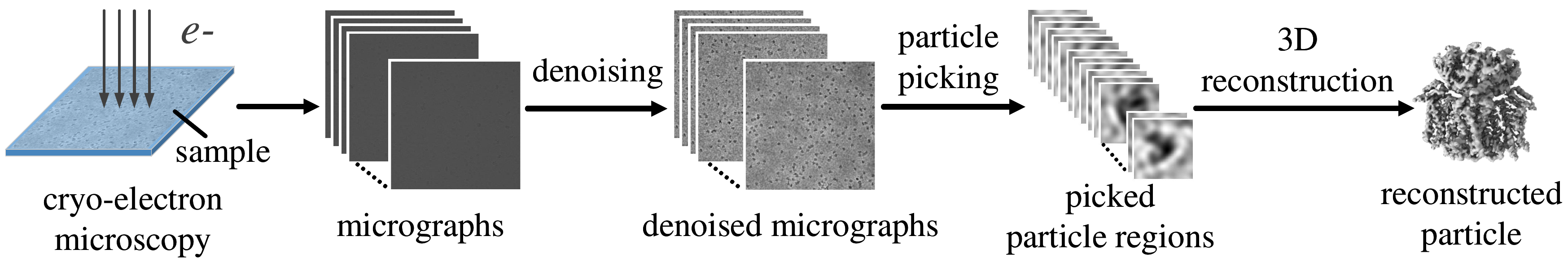}
  \caption{In cryo-EM with SPA, electron beams capture numerous 2D images of proteins within a cryogenically preserved sample. These images are subsequently denoised and subjected to particle picking, facilitating the reconstruction of the 3D structure of the protein. }
  \label{fig:01}
  
\end{figure}

Cryo-EM is vital for obtaining high-resolution images of biological entities, such as cells, viruses, and proteins, at cryogenic temperatures, significantly minimizing radiation damage.  It has revolutionized structural biology, especially through single-particle analysis (SPA), allowing for the detailed examination of molecular structures in their near-native state \cite{milne2013cryo}.
The process starts with sample preparation, where specimens are vitrified in a thin ice layer to maintain their native state. Researchers then use a transmission electron microscope to gather multiple 2D projection images from different angles. Image processing includes denoising and identifying particles for 3D reconstruction. Fig.~\ref{fig:01} presents a simplified workflow of SPA using cryo-EM \cite{zhang2023genem}.

Particle picking is a pivotal step in cryo-EM for isolating individual protein particles from micrographs for further analysis. The quality of particle picking significantly influences the accuracy and resolution of the reconstructed particle structure in the following steps. Challenges in particle picking include the low SNR and varied particle orientations in cryo-EM micrographs, necessitating a large sample size for accurate 3D reconstructions \cite{bepler2019positive}. Moreover, manual picking is inefficient, time-consuming, labor-intensive, error-prone, and introduces dataset inconsistencies \cite{dhakal2023large}. Mis-identifications, or false positives, further compromise reconstruction quality. These issues highlight the need for improved particle selection techniques to enhance both the efficiency of particle identification and the overall quality of cryo-EM reconstructions, emphasizing the reduction of false positives and the increase of true positives \cite{li2022transfer}. 

Various semi-automated and automated cryo-EM particle picking methods have been developed in response to this need. Traditional methods are categorized into template-free \cite{pei2016simulating} and template-based methods \cite{punjani2017cryosparc, rigort2012automated,scheres2015semi, tang2007eman2}.
Template-free methods like the Difference of Gaussians (DoG) \cite{voss2009dog} are noise-sensitive and less effective for irregular particles. Template-based approaches struggle with particle variability and are ill-suited for novel structures, limiting their efficacy in complex cryo-EM analysis. With the advent of deep learning, NN-based particle picking methods \cite{bepler2019positive,wagner2019sphire,wang2016deeppicker,zhu2017deep} have been proposed, marking a significant evolution in the field. These advanced techniques leverage the powerful pattern recognition capabilities of deep learning models to enhance the accuracy and efficiency of particle picking. 
Among these methods, crYOLO \cite{wagner2019sphire} and Topaz \cite{bepler2019positive} are notable for their widespread application. While crYOLO is recognized for its efficiency in particle detection, it occasionally misses real particles. Topaz, though capable of identifying particles with limited labeled data, is susceptible to false positives and duplicates. Despite claims of minimal data requirements, these methods 
still often require large-scale labeled datasets for improved performance. Moreover, they exhibit limited generalization to unseen data, restricting their applicability in diverse cryo-EM research settings.

In this study, we present \textbf{cryoMAE}, a cutting-edge cryo-EM particle picking approach, drawing inspiration from MAE \cite{he2022masked}. Leveraging the few-shot learning paradigm, cryoMAE is meticulously designed to first learn representative particle features from a limited set of cryo-EM particle regions efficiently, cryoMAE then detects and extracts particles from query micrographs by comparing the latent features generated for exemplars against those from regions within the query micrographs.
The operation of cryoMAE unfolds in two distinct stages.  Initially, it trains on a curated set of particle regions and a broader selection of unlabeled regions from a reference micrograph, utilizing a self-supervised approach. We introduce a unique self-cross similarity loss, ensuring the cryoMAE encoder generates distinct latent features for particle and non-particle areas. Subsequently, the trained encoder analyzes query micrographs, extracting and comparing latent features to exemplar features to ascertain particle locations through similarity scoring. 

The performance of cryoMAE was rigorously evaluated using the CryoPPP cryo-EM particle picking dataset \cite{dhakal2023large}, showcasing significant enhancements in 3D particle reconstruction resolution. Particles selected by our model from this dataset exhibit up to $22.4\%$ (average $11.1\%$) improvement in resolution compared to those picked using current SOTA models. Remarkably, these results were achieved using just a few labeled exemplars (\textit{e.g.}, 15) per protein type, highlighting cryoMAE's efficient use of limited data. 

Our contributions are summarized as follows:
\begin{enumerate}
    \item We introduce CryoMAE, an innovative two-stage few-shot learning method specifically designed for SPA cryo-EM particle picking task. This approach markedly diminishes the reliance on extensive, labeled datasets.
    \item We propose a novel formulation of self-cross similarity loss, aiming to promote the model capacity to differentiate between particle objects and background regions. 
    \item Our experimental findings indicate that cryoMAE achieves up to $22.4\%$ improvement in the resolution of 3D particle reconstructions compared to SOTA NN-based particle picking methods.
\end{enumerate}

\section{Related Work}
\label{sec:relate}
\subsubsection{Particle Picking.}
Introduced in Section~\ref{sec:intro}, a variety of approaches exist for cryo-EM particle picking, ranging from automated to semi-automated techniques. Template matching, use predefined reference images or ``templates'' and cross-correlation for particle identification \cite{vinzenz2012actin,zeng2023high}, performing best with known particle structures but limited by template quality and diversity. In contrast, template-free methods bypass the need for templates, employing computer vision techniques to distinguish particles. For instance, the DoG method emphasizes particles by contrasting two differently blurred image versions, improving visibility but risking noise amplification in low-SNR scenarios.

NN-based particle picking methods \cite{bepler2019positive,wagner2019sphire,wang2016deeppicker,zhu2017deep} provide significant advances in cryo-EM, offering more accurate, efficient, and accessible solutions. These methods can learn from a wide range of particle shapes, sizes, and orientations directly from the training data, making them more adaptable to different datasets without the need for specific templates.
CrYOLO \cite{wagner2019sphire} and Topaz \cite{bepler2019positive} are distinguished for their advanced particle picking capabilities in cryo-EM. CrYOLO leverages the You Only Look Once framework \cite{redmon2016you} for particle detection, and Topaz employs convolutional neural networks (CNNs) with positive-unlabeled (PU) learning. Despite their strengths, crYOLO may overlook true particles, while Topaz is prone to recognizing numerous false positives and duplicates \cite{gyawaliaccurate2023}.  They require extensive labeled datasets, demanding significant time and resources. Our cryoMAE, utilizing few-shot learning, offers high efficiency using a minimal number of exemplars. It effectively reduces false negatives and positives, and minimizes reliance on large labeled datasets, representing a significant leap in cryo-EM particle picking technology.

\subsubsection{Masked Autoencoders.}
MAEs were initially introduced by He et al. \cite{he2022masked}, drawing inspiration from the BERT model \cite{devlin2018bert}, a transformative approach in natural language processing. MAEs bring the innovative concept of masking into the realm of computer vision, a technique where random sections of an image are obscured (masked) before being processed by an encoder. Subsequently, a decoder attempts to reconstruct these masked sections. \cite{he2022masked} demonstrated that masking a substantial portion of the image (up to $75\%$) compels the model to learn deeper and more comprehensive representations of the data. In our study, we harness the exceptional feature extraction capabilities of MAEs to discern unique features of particles, thereby enhancing the efficiency and accuracy of particle picking in cryo-EM.

\subsubsection{Contrastive Learning.}
Contrastive Learning has been a transformative force in unsupervised learning, concentrating on increasing the similarity between representations of positive pairs while simultaneously differentiating those of negative pairs. Pioneering this approach, the concept of contrastive loss was introduced for dimensionality reduction and embedding learning, aiming to preserve semantic similarity \cite{hadsell2006dimensionality}. Further advances have been made with the development of SimCLR \cite{chen2020simple}, which utilizes data augmentation techniques to enhance the robustness of visual representations. Moreover, He et al. \cite{he2020momentum} introduced Momentum Contrast, a methodology for building dynamic dictionaries in contrastive learning, which refines the application of contrastive loss. This refinement ensures the consistency of the representations for negative samples across the learning process. In our research, we leverage the principles of contrastive learning to develop a unique contrastive loss mechanism called self-cross similarity loss. This innovation enables our model to effectively discriminate between regions containing particles and background regions.

\section{Methodology}
In this section, we detail cryoMAE, starting with defining the few-shot cryo-EM particle picking problem, followed by our two-stage framework.
\subsection{Overview}
\subsubsection{Problem setup.}

Given a reference micrograph $\mathbf{R}$, containing the target particles for analysis, we first randomly select a reference micrograph $\mathbf{R}$ and manually label $m$ ($m$ is a small number, \textit{e.g.} 15) particle regions $\mathbf{x}_i^l$ as exemplars ($\textbf{X}_\textbf{L} = \{\mathbf{x}_i^l\}_{i=1}^{m}$), and randomly crop additional $n$ regions $\mathbf{x}_j^u$ from the same cryo-EM micrograph as unlabeled regions ($\textbf{X}_\textbf{U} = \{\mathbf{x}_j^u\}_{j=1}^{n}$). The remaining micrographs containing the same particle are query micrograph set $\mathbf{Q}$.
Our goal is to leverage the limited set of exemplars $\mathbf{X}_\mathbf{L}$ and unlabeled regions extracted from $\mathbf{R}$ to detect the particles within $\mathbf{R}$ and $\mathbf{Q}$.

\begin{figure}[!htb]
  \centering
  \includegraphics[width=0.8\textwidth]{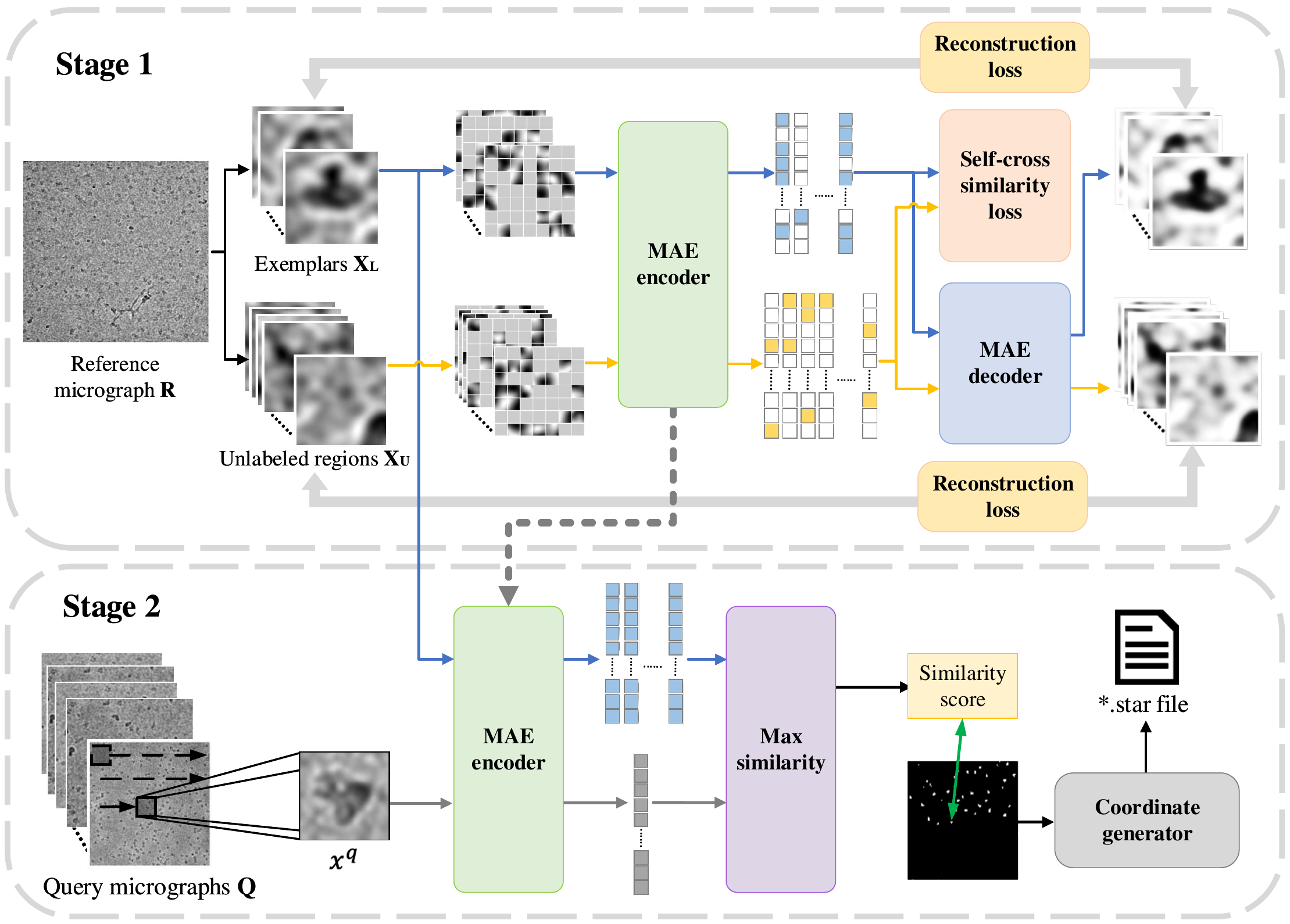}
  \caption{Overview of the two-stage cryoMAE framework: stage 1 illustrates the training phase with a mix of labeled particle and unlabeled regions, employing reconstruction loss and self-cross similarity loss. Stage 2 depicts the particle picking process, where the trained MAE encoder assesses query micrographs, leveraging latent feature comparisons to identify particle positions accurately.
  }
  \label{fig:02}
\end{figure}

As depicted in Fig.~\ref{fig:02}, our framework unfolds in two distinct stages. In stage 1, cryoMAE is trained using a mixture of labeled exemplars $\mathbf{X}_\mathbf{L}$ and unlabeled regions $\mathbf{X}_\mathbf{U}$ from $\mathbf{R}$. This training process is guided by both mean squared error reconstruction loss and a novel self-cross similarity loss, which helps the model distinguish between regions with and without particles. In stage 2, trained MAE encoder scans query micrographs to identify particles, comparing latent features of regions against those of exemplars to determine similarity scores. Regions with higher similarity scores are identified as more likely to contain particles, facilitating accurate particle picking.

\subsection{Stage 1: Training on One Reference Micrograph}
\label{sec:stage1}
\subsubsection{Model training.}
For each protein type represented by multiple micrographs, we select a reference micrograph $\mathbf{R}$ with manually annotated regions $\mathbf{X}_\mathbf{L}$ as exemplars and crop random unlabeled regions $\mathbf{X}_\mathbf{U}$ from the remaining parts of $\mathbf{R}$. As discussed in \cite{bepler2019positive}, particle regions are sparse within micrographs, making most unlabeled regions likely non-particle areas. These images are resized to $224 \times 224$ and further processed into $16 \times 16$ patches during training, which are then subjected to random masking at a rate of $75\%$. This process transforms exemplar and unlabeled regions into $\hat{\mathbf{x}}_i^l$ for labeled exemplars and $\hat{\mathbf{x}}_j^u$ for unlabeled regions, respectively.  The cryoMAE encoder then generates latent features for these regions, denoted as $\mathbf{E}(\hat{\mathbf{x}}_i^l)$ and $\mathbf{E}(\hat{\mathbf{x}}_j^u)$ respectively. Subsequently, the MAE decoder utilizes the generated latent features to reconstruct the original input images. This reconstruction is achieved through a self-supervised process, with the original images serving as the supervisory signal.
This masking encourages the model to focus on global features of cryo-EM images, enhancing understanding of particle structures and generalizing across conditions. Such a focus is crucial for overcoming the limited training data challenge in the cryo-EM field, improving the model's performance in particle detection and generalization.

Training cryoMAE incorporates both particle and unlabeled regions to bolster model robustness. Exclusive training on particle images could lead MAE to converge towards a homogeneous latent feature space for any given input, potentially escalating the false positive rate by assigning high similarity scores indiscriminately, including to background regions. By including unlabeled regions, cryoMAE learns to recognize features of non-particle spaces, avoiding overfitting to a solely particle-focused feature space. This broader training approach refines the model's ability to distinguish between particle and background regions, markedly lowering false positive rates by assigning more accurate similarity scores to non-particle areas. However, adding unlabeled regions faces some challenges: 1) the diverse background noise in cryo-EM, ranging from crystalline ice contamination and malformed particles to grayscale background regions, which demands a nuanced approach for accurate differentiation; 2) merely incorporating unlabeled data might not prompt the model to learn features unique to particles against complex backgrounds. To optimize the training efficiency of cryoMAE few-shot particle datasets and reduce overfitting risks, while also accounting for a wide range of background noise, we introduced a pre-training phase. Pre-training cryoMAE on a broader set of unlabeled regions better represents background variability. Further, introducing a self-cross similarity loss specifically addresses these noise issues, enhancing the model's ability to discern particles from backgrounds.

\subsubsection{Self-cross similarity.}
\begin{figure}[!htb]
  \centering
  \includegraphics[width=0.425\textwidth]{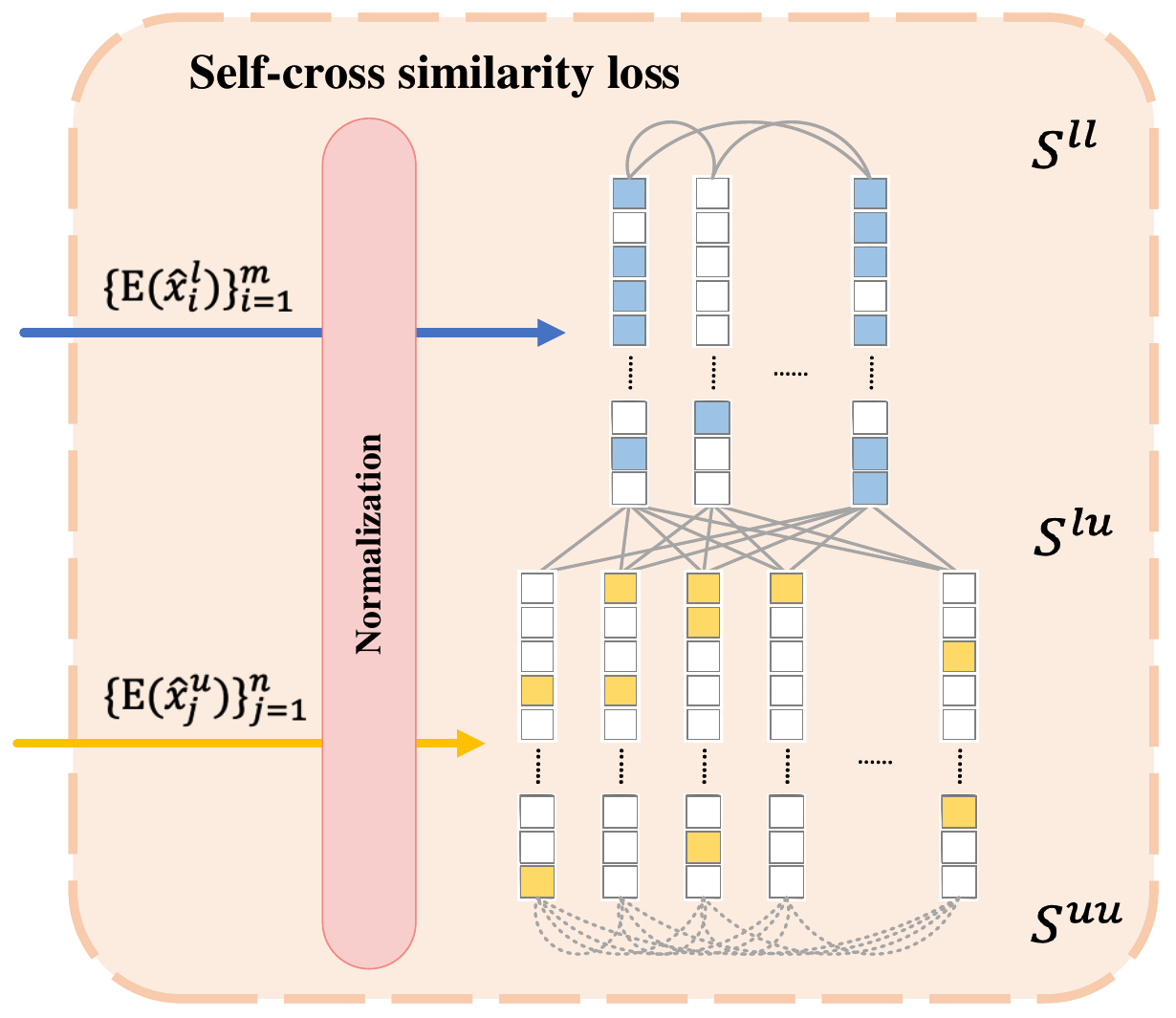}
  \caption{Self-cross similarity loss.}
  \label{fig:scs-framework}
\end{figure}

Drawing from the self-similarity concept \cite{shi2022represent}, we develop a self-cross similarity loss to foster distinct latent features for particles and background within cryo-EM images, enhancing the model's ability to differentiate between these regions. This approach aims to increase the disparity in the feature space, thereby improving the precision of particle identification. As illustrated in Fig.~\ref{fig:02}, the MAE encoder's latent features are utilized not only for image reconstruction by the decoder but also are evaluated using the self-cross similarity loss, further detailed in Fig.~\ref{fig:scs-framework}. The cosine similarity between two feature vectors $\mathbf{a}$ and $\mathbf{b}$ is calculated as $S_{cos}(\mathbf{a}, \mathbf{b}) = \frac{\mathbf{a}^T \mathbf{b}}{ \|\mathbf{a}\| \|\mathbf{b}\|}$.

The self-similarity $S_{self}$ is calculated as the mean cosine similarity among the features of positive regions, formalized as:
\begin{equation}
  S_{self} = \frac{1}{m^2} \sum\nolimits_{i=1}^{m} \sum\nolimits_{j=1}^{m}  S_{cos}(\text{E}(\hat{\mathbf{x}}_i^l), \text{E}(\hat{\mathbf{x}}_j^l)).
\end{equation}
Similarly, the cross similarity $S_{cross}$ is the mean cosine similarity between features of positive and unlabeled regions:
\begin{equation}
  S_{cross} = \frac{1}{mn} \sum\nolimits_{i=1}^{m} \sum\nolimits_{j=1}^{n} S_{cos}(\text{E}(\hat{\mathbf{x}}_i^l), \text{E}(\hat{\mathbf{x}}_j^u)).
\end{equation}
\(S_{self}\) measures the similarity among latent features of exemplars, reflecting the internal consistency of particle features. This is crucial for the model to identify and enhance particle-specific patterns, facilitating better distinction from background noise. Ideally, The goal is for \(S_{self}\) to increase, indicating stronger similarity within particle groups. Conversely, \(S_{cross}\) assesses the similarity between exemplar features and those of unlabeled (negative) regions, aiming to capture the distinctiveness between particles and background. The objective is for \(S_{cross}\) to decrease, signifying reduced feature similarity between particles and background. Self-cross similarity loss \(L_{SCS}\) is designed to optimize these dynamics, thereby improving model's ability to differentiate between particles and backgrounds:
\begin{equation}
  L_{SCS}(S_{cross}, S_{self}) = \max\left(\tau, 1 + \alpha \cdot S_{cross} - (1 - \alpha) \cdot S_{self}\right).
  \label{eq:important}
\end{equation}
\(\alpha\) balances self and cross-similarity contributions, and \(\tau\) sets a minimum difference threshold between them, limiting further distinction efforts beyond it.

\subsubsection{PU learning.}
Inspired by \cite{bepler2019positive}, we identify a limitation in our previous loss function design, which treats unlabeled data as negative. Randomly cropped training regions may unintentionally include particles, potentially confusing the model's distinction between labeled particles and background noise. This overlap complicates training, as the model could wrongly link particle features with the background, undermining our strategy to reduce background similarity scores and challenging the model's ability to learn discriminatively. 
To enhance the loss formulation, we accommodate the potential inclusion of particles in unlabeled regions. Acknowledging that a certain proportion ($\hat{\pi}$) of these samples may harbor particles, we modify the representation of features for these unlabeled samples.

We adjust feature representation by implementing a weighting scheme grounded in the estimated probability $\hat{\pi}$ that an unlabeled region harbors a particle, alongside a complementary weight $1-\hat{\pi}$ for regions likely devoid of particles. This probabilistic approach enhances the model's capacity to differentiate between particle-laden regions and pure background, optimizing the use of unlabeled data in training and improving particle identification accuracy.
The presence of particles in unlabeled regions necessitates a recalibration of similarity calculations, introducing a deeper analysis of self-similarity among potential positives and their cross-similarity with potential negatives within the unlabeled data:

\begin{equation}
\begin{split}
\hat{S}_{self} = \frac{1}{(m+n)^2} \bigg[ & S^{ll} + 2\hat{\pi} S^{lu} + \hat{\pi}^2 S^{uu} \bigg],\\
\end{split}
\label{eq:self}
\end{equation}
\begin{equation}
\begin{split}
\hat{S}_{cross} = \frac{1}{(m+n)\times n} \bigg[ &  (1-\hat{\pi}) S^{lu} + \hat{\pi}(1-\hat{\pi}) S^{uu} \bigg]. \\
\end{split}
\label{eq:cross}
\end{equation}
$S^{ll}$, $S^{lu}$, and $S^{uu}$ measure the sums of cosine similarities among exemplars, between exemplars and unlabeled regions, and among unlabeled regions, respectively. In the formulas, we decide not to adjust $n$ because we treat each latent feature adjustment as a weighted process. Under this logic, we view it as having $n$ latent features adjusted by $\hat{\pi}$ and $1-\hat{\pi}$, rather than having a total of $\hat{\pi}n$ particle regions or $(1-\hat{\pi})n$ background regions within all unlabeled regions. This enhances the clarity of our methodology and ensures its alignment with Fig.~\ref{fig:scs-framework}, thereby preserving logical coherence. The refined self-cross similarity loss, $\hat{L}_{SCS}(\hat{S}_{cross}, \hat{S}_{self})$, adeptly captures the complexity of similarity within data subsets. By refining these calculations, we refine these metrics to account for the intricate characteristics of unlabeled data, facilitating a more discerning and efficacious training regimen.

The total loss of cryoMAE, taking into account the reconstruction loss:
\begin{equation}
  L_{total} = L_{MSE} 
  + \beta \cdot \hat{L}_{SCS}.
  \label{eq:cryoloss}
\end{equation}
Here $\beta$ adjusts the weight of the self-cross similarity loss in the overall loss function, balancing reconstruction accuracy with discriminative learning.

\subsection{Stage 2: Particle Picking on Query Micrographs}
In stage 2, our model undertakes particle picking by utilizing the MAE encoder to scan query micrographs and extract features from each sliding window region, as detailed in \textbf{Stage 2} of Fig.~\ref{fig:02}. This stage does not employ masking for the input regions. The extracted latent features are then matched against those of exemplars through cosine similarity, assigning similarity scores to each region based on the highest similarity. Following the completion of the sliding process on a micrograph, these similarity scores are ranked. It is crucial to recognize the variability in the imaging states of different micrographs, where a single threshold does not work well. Therefore, we adopt a density-based method to determine the most suitable cutoff threshold for each micrograph automatically. This process involves calculating the average distance of each score to its $k$ nearest neighbors, and finding the score where the rate of change in these average distances is maximized as the cutoff threshold. Coordinates of all regions with similarity scores exceeding this threshold, along with the micrograph filenames, are recorded in a .star file. The .star format is widely used in cryo-EM to document particle coordinates, aiding in subsequent steps like 3D reconstruction using CryoSPARC.

\section{Experiments}

This section evaluates cryoMAE against SOTA particle picking methods using the CryoPPP dataset, including ablation studies, sensitive analysis, and qualitative visualizations to demonstrate its effectiveness.

\subsection{Experimental Setup}

\subsubsection{Datasets.}
We evaluated cryoMAE using five distinct particle datasets from CryoPPP \cite{dhakal2023large}, which were obtained from the Electron Microscopy Public Image Archive (EMPIAR) database \cite{iudin2023empiar}. EMPIAR is a publicly accessible resource that offers raw, high-resolution cryo-EM images for research and benchmarking in the field of electron microscopy. The datasets used in our experiments, identified by EMPIAR IDs 10081, 10093, 10345, 10532, and 11056, comprise 300, 300, 300, 300, and 361 micrographs, respectively, each accompanied by particle coordinate information. Each EMPIAR ID corresponds to a unique protein type, facilitating targeted analysis within our SPA framework.

\subsubsection{Baselines.}
In this study, we utilized crYOLO\footnote{\url{https://cryolo.readthedocs.io/en/stable/}.} 
\cite{wagner2019sphire} and Topaz\footnote{\url{https://cb.csail.mit.edu/cb/topaz/}.} 
\cite{bepler2019positive} introduced in Section~\ref{sec:relate} as our baselines. For crYOLO, we employed the general model for crYOLO pre-trained on more than 40 datasets that can select particles of previously unseen macromolecular species as claimed in \cite{wagner2019sphire}. For Topaz, we used a pre-trained model based on ResNet \cite{he2016deep} (16 layers, each layer has 64 units) trained on large-scale cryo-EM datasets. 

\subsubsection{Evaluation metrics.}
Our evaluation metrics include precision, recall, and F1 scores. A true positive occurs when a picked particle region overlaps with a ground truth region, achieving an intersection over union (IoU) of 0.5 or higher, with each ground truth accounted for only once. False positives include picked regions that either have an IoU less than 0.5 with any ground truth region or represent multiple detections for a single ground truth. False negatives are ground truth regions that remain undetected.

\subsubsection{Particle picking.}
The cryoMAE encoder slides on and processes query images in stage 2 with a stride of 28, extracting features for each sub-region. These features are matched against exemplar features, assigning the highest similarity score to each region. Following the sliding process, scores are ordered, and a density-based approach determines the cut-off threshold by identifying a sharp change in the 5 nearest neighbor average distance list. Coordinates from regions above this threshold are pinpointed as particle locations.

\subsubsection{3D reconstruction.}
We utilized CryoSPARC \cite{punjani2017cryosparc} to conduct 3D reconstructions on particles selected by various methods and compared the resolutions of the reconstructed particles. The workflow, from particle picking to reconstructed structure, encompasses essential steps: contrast transfer function (CTF) estimation, 2D classification, 2D class selection, \textit{ab initio} reconstruction, and homogeneous refinement. CTF estimation corrects for the microscope's phase contrast, crucial for high-resolution reconstructions. 2D classification sorts particles into classes, removing aberrant particles to improve data quality. 2D class selection further ensures only high-quality particles are used, followed by \textit{ab initio} reconstruction for an initial 3D model creation without prior knowledge.

\subsection{Overall Performance}

\begin{table}[!htb]
  \caption{Performance comparison of cryoMAE, crYOLO, and Topaz on CryoPPP.}
  \label{tab:comparison1}
  \centering
  \resizebox{\textwidth}{!}{
  \begin{tabular}{@{}cccccccccccccccc@{}}
    \toprule
      \multirow{3}{*}{EMPIAR ID} & \multicolumn{2}{c}{Data information} & \multicolumn{3}{c}{Precision} & \multicolumn{3}{c}{Recall} & \multicolumn{3}{c}{F1 score} \\
       \cmidrule(lr){2-3} \cmidrule(lr){4-6} \cmidrule(lr){7-9} \cmidrule(lr){10-12}
       & \multirow{2}{*}{Image size} & \multicolumn{1}{c}{Particle} & \multirow{2}{*}{crYOLO} & \multirow{2}{*}{Topaz} & \multirow{2}{*}{Ours} & \multirow{2}{*}{crYOLO} & \multirow{2}{*}{Topaz} & \multirow{2}{*}{Ours} & \multirow{2}{*}{crYOLO} & \multirow{2}{*}{Topaz} & \multirow{2}{*}{Ours} \\
       & & \multicolumn{1}{c}{diameter (px)} & & & & & & & & & \\
      \midrule
    10081 & (3710, 3838)& 154& \textbf{0.705} & 0.412 & {0.645} & {0.867} & 0.855 & \textbf{0.939} & \textbf{0.777} & 0.556 & {0.765} \\
    10093 & (3838, 3710)& 172& {0.380} & 0.328 & \textbf{0.383} & {0.355} & 0.209 & \textbf{0.497} & {0.367} & 0.255 & \textbf{0.433} \\
    10345 & (3838, 3710)& 149& {0.441} & 0.195 & \textbf{0.473} & 0.561 & {0.732} & \textbf{0.733} & {0.494} & 0.308 & \textbf{0.575} \\
    10532 & (4096, 4096)&174 & {0.501} & 0.387 & \textbf{0.503} & 0.231 & {0.311} & \textbf{0.497} & 0.316 & {0.345} & \textbf{0.500} \\
    11056 & (5760, 4092)&164 & {0.690} & 0.453 & \textbf{0.694} & 0.465 & {0.578} & \textbf{0.671} & {0.556} & 0.507 & \textbf{0.682} \\
    \midrule
    Average   & -&- &\textbf{0.543} & 0.355 & {0.540} & 0.496 & {0.537} & \textbf{0.667} & {0.502} & 0.394 & \textbf{0.591} \\
    \bottomrule
  \end{tabular}}
\end{table}

\begin{table}[!htb]
  \caption{\textit{Ab-initio} reconstruction resolution comparison of cryoMAE, crYOLO, and Topaz across EMPIAR Datasets from CryoPPP. 
  }
  \label{tab:comparison2}
  \centering\resizebox{0.8\textwidth}{!}{
  \begin{tabular}{@{}ccccccccccccc@{}}
    \toprule
    \multirow{2}{*}{EMPIAR ID}  & Protein & \# of & \# of & \multicolumn{3}{c}{Resolution (Å)} \\
    \cmidrule(l){5-7}
      & type& micrographs& GT particles&crYOLO & Topaz & Ours  \\
    \midrule
    10081 &Transport protein& 300 &39,352 & 12.25 & 12.72 & \textbf{11.32} \\
    10093 &Membrane protein& 300 &56,394 & 11.64 & 11.62 & \textbf{9.02} \\
    10345 &Signaling protein& 300 &15,894 & 11.63 & 10.39 & \textbf{10.27} \\
    10532 &Viral protein& 300 &87,933 & 12.86 & 10.85 & \textbf{9.92} \\
    \midrule
    Average   &-& - & - & 12.10 & 11.40 & \textbf{10.13}\\
    \bottomrule
  \end{tabular}}
  
\end{table}

\begin{figure}[!htb]
  \centering

  \includegraphics[width=\textwidth]{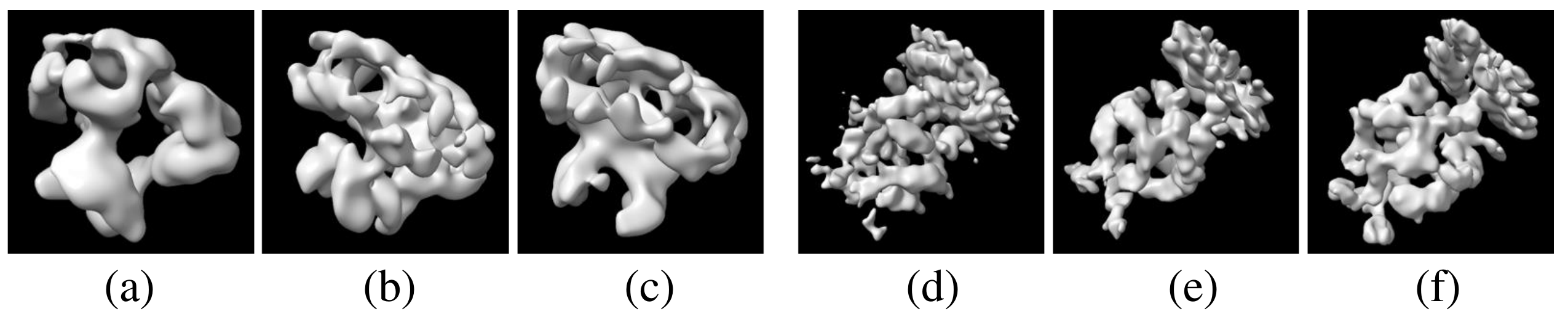}
  \caption{3D reconstructions for EMPIAR-10081 and EMPIAR-10093 using crYOLO, Topaz, and cryoMAE: (a)-(c) for 10081, (d)-(f) for 10093.
  }

  \label{fig:3dreconstruction}
  
\end{figure}

The performance comparison of crYOLO, Topaz, and cryoMAE in particle picking is detailed in Tables~\ref{tab:comparison1} and \ref{tab:comparison2}, with 3D reconstruction outcomes visualized in Fig.~\ref{fig:3dreconstruction}.  Table~\ref{tab:comparison1} reveals the high precision of crYOLO but its tendency to miss many particles. This is particularly evident with EMPIAR-10081, where crYOLO shows strong performance due to pre-training on a dataset that included EMPIAR-10081 particles. This pre-trained crYOLO model raises questions about its generalization to new particle types, where performance significantly drops, highlighting a generalization issue.
Topaz scores well in the recall, albeit with a high false positive rate. Conversely, cryoMAE excels in both precision and recall, outperforming Topaz in all evaluated metrics for five particles and showing better recall than crYOLO, aside from EMPIAR-10081. It surpasses crYOLO in precision and F1 score, excluding EMPIAR-10081. Importantly, cryoMAE leads to the highest 3D reconstruction resolution on CryoPPP dataset particles, indicating an average $11\%$ resolution improvement over baselines.

\subsection{Ablation Studies}
Ablation studies validate the contributions of key cryoMAE components: self-cross similarity loss, pre-training, and exemplar similarity matching.

\begin{table}[!htb]
  \caption{Comparison of cryoMAE supervised w/o self-cross similarity loss, w/ unadjusted self-cross similarity loss $L_{SCS}$, and w/ adjusted self-cross similarity loss $\hat{L}_{SCS}$.}
  \label{tab:scs}
  \centering
  \resizebox{\textwidth}{!}{
  \begin{tabular}{@{}cccccccccc@{}}
    \toprule
    \multirow{2}{*}{EMPIAR ID} & \multicolumn{3}{c}{Precision} & \multicolumn{3}{c}{Recall} & \multicolumn{3}{c}{F1 Score} \\
    \cmidrule(l){2-4} \cmidrule(l){5-7} \cmidrule(l){8-10}
      & w/o & w/ $L_{SCS}$ & w/ $\hat{L}_{SCS}$ & w/o & w/ $L_{SCS}$ & w/ $\hat{L}_{SCS}$ & w/o & w/ $L_{SCS}$ & w/ $\hat{L}_{SCS}$ \\
    \midrule
    10081 & 0.225 & 0.639 & \textbf{0.645} & 0.652 & 0.928 & \textbf{0.939} & 0.335 & 0.757 & \textbf{0.765} \\
    10093 & 0.143 & 0.376 & \textbf{0.383} & 0.216 & 0.493 & \textbf{0.497} & 0.172 & 0.427 & \textbf{0.433} \\
    10345 & 0.180 & \textbf{0.474} & 0.473 & 0.547 & 0.724 & \textbf{0.733} & 0.271 & 0.573 & \textbf{0.575} \\
    10532 & 0.177 & 0.495 & \textbf{0.503} & 0.269 & 0.478 & \textbf{0.497} & 0.213 & 0.486 & \textbf{0.500} \\
    11056 & 0.154 & 0.691 & \textbf{0.694} & 0.327 & 0.639 & \textbf{0.671} & 0.209 & 0.664 & \textbf{0.682} \\
    \midrule
    Average   & 0.176 & 0.535 & \textbf{0.540} & 0.402 & 0.652 & \textbf{0.667} & 0.240 & 0.581 & \textbf{0.591} \\
    \bottomrule
  \end{tabular}}
\end{table}

\begin{figure}[!htb]
  \centering
  \begin{minipage}{.46\textwidth}
  \centering
  \includegraphics[width=\textwidth]{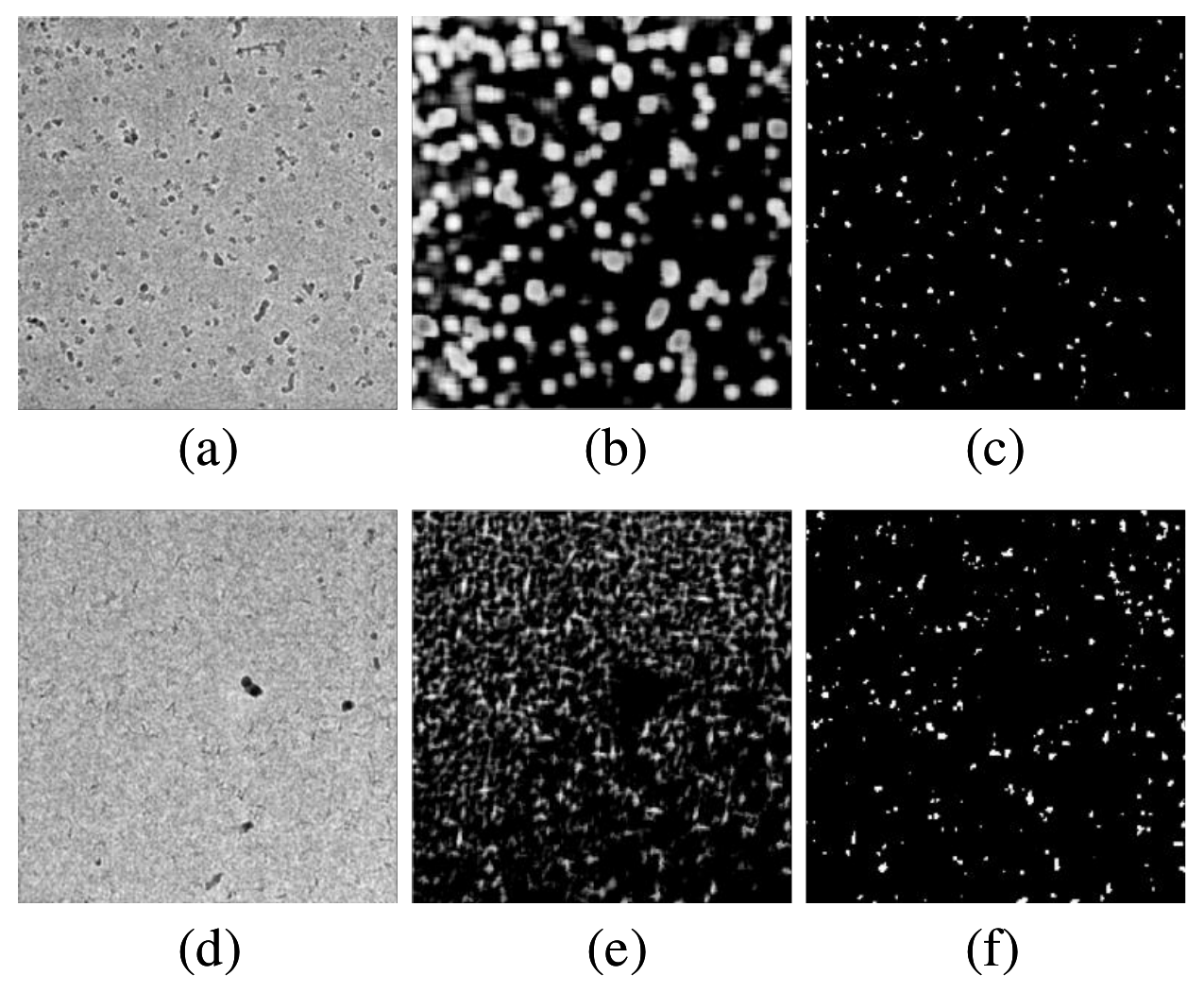}
  \caption{Similarity maps generated from query micrographs by cryoMAE, w/ and w/o adjusted self-cross similarity loss. (a),(d) original micrographs; (b),(e) similarity map w/o adjusted self-cross similarity loss; (c),(f) w/ adjusted self-cross similarity loss.
  }
  \label{fig:scscompare}
  \end{minipage}%
  \hfill
  \begin{minipage}{.46\textwidth}
  \vspace{-0.24cm}
  \centering
  \includegraphics[width=0.95\textwidth]{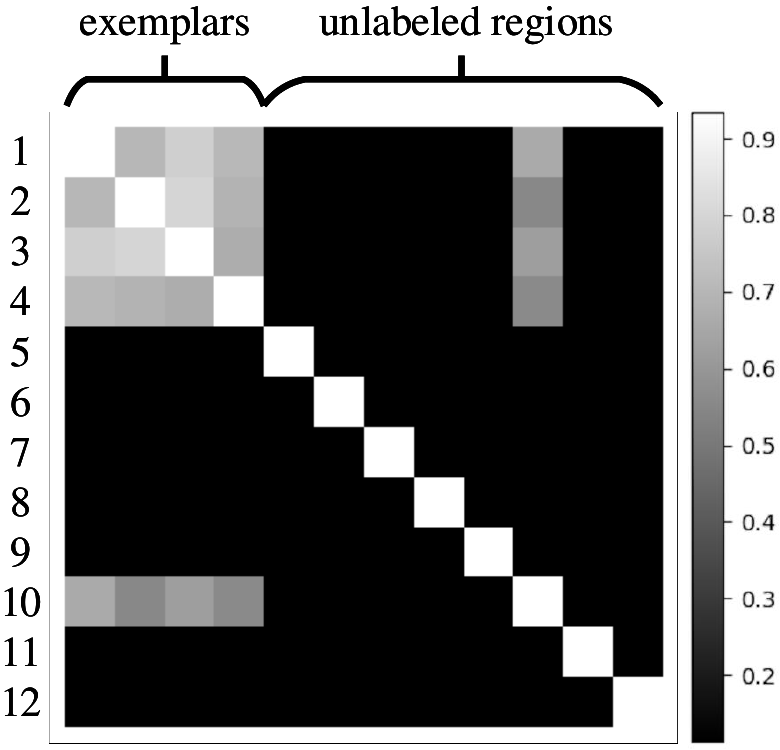}
  \caption{Cosine similarity matrix for 12 regions, comprising 4 exemplars (1-4) and 8 unlabeled regions (5-12). Entries at the intersection of row $i$ and column $j$ denote the cosine similarity between latent features of regions $i$ and $j$.
  }
  \label{fig:scscompare2}
  \end{minipage}
\end{figure}

\begin{figure}[tb]
  \centering
  \begin{subfigure}{0.32\linewidth}
    \includegraphics[width=\textwidth]{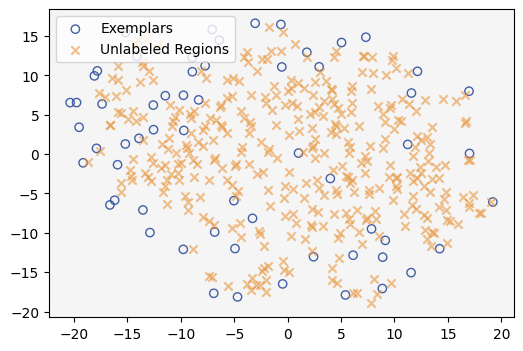}
    \caption{trained on a dataset w/o unlabeled regions.}
    \label{fig:short-a}
  \end{subfigure}
  \hfill
    \begin{subfigure}{0.32\linewidth}
    \includegraphics[width=\textwidth]{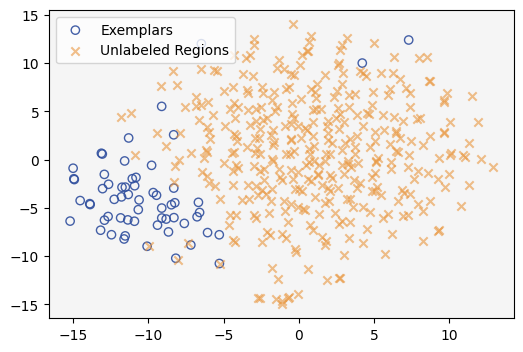}
    \caption{trained w/o the adjusted self-cross similarity loss.}
    \label{fig:short-b}
  \end{subfigure}
  \hfill
  \begin{subfigure}{0.32\linewidth}
    \includegraphics[width=\textwidth]{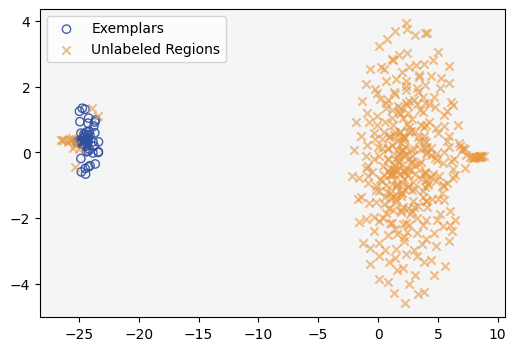}
    \caption{trained w/ the adjusted self-cross similarity loss.}
    \label{fig:short-c}
  \end{subfigure}
  \caption{2D t-SNE visualizations of cryoMAE latent feature space (on EMPIAR-10081).}
  \label{fig:2dvis}
\end{figure}

\subsubsection{W/ and w/o self-cross similarity loss.}
We assessed the performance of cryoMAE across different configurations of self-cross similarity loss (with self-cross similarity loss, with unadjusted self-cross similarity loss, and with adjusted self-cross similarity loss) in Table~\ref{tab:scs}, revealing optimal performance with adjusted loss. This finding highlights the crucial impact of self-cross similarity loss in enhancing feature extraction, making cryoMAE more discerning in particle selection and greatly lowering the chance of incorrect region identification.
CryoMAE without self-cross similarity loss incorrectly scores many non-particle regions highly, evident from widespread white areas in Fig.\ref{fig:scscompare}(b)(e). In contrast, with this loss, cryoMAE specificity improves, accurately identifying particle regions, as shown in Fig.\ref{fig:scscompare}(c)(f), reducing false scores for background areas.
Further insights are shown in Fig.~\ref{fig:scscompare2}, displaying a cosine similarity matrix for 12 regions, including 4 exemplars (1-4) and 8 unlabeled areas (5-12), with region 10 being a particle region. The matrix demonstrates high similarity among particle regions and lower similarity between particle and background regions, highlighting the model's ability to group particle regions closely in the feature space and distinguish them from the background. This is key to the success of the self-cross similarity loss, enabling the model to significantly reduce similarity scores for non-target areas and concentrate high scores on central particle regions, thus reducing false positives. Conversely, models trained without this loss struggle to separate particle regions from backgrounds, leading to increased false positives.

We also conduct 2D t-SNE visualizations to analyze the latent features of cryoMAE under varying conditions: trained on a dataset without unlabeled regions, trained on a dataset with unlabeled regions without the adjusted self-cross similarity loss, and trained on a dataset with unlabeled regions with the adjusted self-cross similarity loss. For each visualization, we randomly select a consistent set of 60 exemplars and 360 unlabeled regions from EMPIAR-10081 to ensure comparability across the three scenarios. The visualizations are in Fig.~\ref{fig:short-a}, Fig.~\ref{fig:short-b} and Fig.~\ref{fig:short-c}, respectively. As demonstrated in Fig.~\ref{fig:short-a}, training exclusively on particle regions leads cryoMAE to generate homogeneous latent features for any input. This approach risks elevating the false positive rate by indiscriminately assigning high similarity scores, including to background regions. Fig.~\ref{fig:short-b} illustrates that incorporating unlabeled regions enables cryoMAE to discern features of non-particle regions, thus mitigating over-fitting to a particle-exclusive feature space. Consequently, the model acquires a preliminary capability to differentiate between particle and background regions, although with limited clarity (as observed in the latent feature space 2D visualization, where the blue and yellow clusters are approximately but not distinctly separated). Further advancements are evident in Fig.~\ref{fig:short-c}, where the introduction of adjusted self-cross similarity loss significantly enhances the model's ability to distinguish between background regions and particles. This improvement is illustrated by the distinct separation between the two clusters in the figure, despite the presence of some yellow points within the blue cluster. These exceptions, representing particle-containing regions within unlabeled areas, are considered reasonable.

\begin{table}[!htb]
    \centering
    \begin{minipage}{.48\linewidth} 
      \caption{W/ and w/o pre-training.}
      \label{tab:pretrain}
      \centering
      
      \resizebox{\textwidth}{!}{\begin{tabular}{@{}cccccccccc@{}}
        \toprule
        EMPIAR & \multicolumn{2}{c}{Precision} & \multicolumn{2}{c}{Recall} & \multicolumn{2}{c}{F1 Score} \\
        \cmidrule(l){2-3} \cmidrule(l){4-5} \cmidrule(l){6-7}
         ID & w/ & w/o & w/ & w/o & w/ & w/o \\
        \midrule
        10081 & \textbf{0.645} & 0.352 & \textbf{0.939} & 0.919 & \textbf{0.765} & 0.509 \\
        10093 & \textbf{0.383} & 0.281 & \textbf{0.497} & 0.479 & \textbf{0.433} & 0.354 \\
        10345 & \textbf{0.473} & 0.209 & \textbf{0.733} & 0.581 & \textbf{0.575} & 0.307 \\
        10532 & \textbf{0.503} & 0.404 & \textbf{0.497} & 0.347 & \textbf{0.500} & 0.373 \\
        11056 & \textbf{0.694} & 0.664 & \textbf{0.671} & 0.579 & \textbf{0.682} & 0.619 \\
        \midrule
        Average   & \textbf{0.540} & 0.382 & \textbf{0.667} & 0.581 & \textbf{0.591} & 0.432 \\
        \bottomrule
      \end{tabular}}
      
    \end{minipage}%
    \hfill
    \begin{minipage}{.50\linewidth} 
  \caption{Max and mean matching. }
  \label{tab:matchingmethod}
      \centering
  \resizebox{\textwidth}{!}{\begin{tabular}{@{}cccccccccc@{}}
    \toprule
    EMPIAR & \multicolumn{2}{c}{Precision} & \multicolumn{2}{c}{Recall} & \multicolumn{2}{c}{F1 Score} \\
    \cmidrule(l){2-3} \cmidrule(l){4-5} \cmidrule(l){6-7}
      ID& Max & Mean & Max & Mean & Max & Mean \\
    \midrule
    10081 & \textbf{0.645} & 0.595 & 0.939 & \textbf{0.946} & \textbf{0.765} & 0.731 \\
    10093 & \textbf{0.383} & 0.367 & 0.497 & \textbf{0.548} & 0.433 & \textbf{0.440} \\
    10345 & \textbf{0.473} & 0.396 & \textbf{0.733} & 0.718 & \textbf{0.575} & 0.510 \\
    10532 & \textbf{0.503} & 0.498 & 0.497 & \textbf{0.502} & \textbf{0.500} & 0.500 \\
    11056 & \textbf{0.694} & 0.606 & \textbf{0.671} & 0.650 & \textbf{0.682} & 0.628 \\
    \midrule
    Average   & \textbf{0.540} & 0.492 & 0.667 & \textbf{0.673} & \textbf{0.591} & 0.562 \\
    \bottomrule
  \end{tabular}}
    \end{minipage} 
    
\end{table}

\subsubsection{W/ and w/o pre-training.}
Table~\ref{tab:pretrain} demonstrates how employing pre-training strategy on cryoMAE markedly promotes model performance, with gains in precision and recall at $41.4\%$, and in F1 score at $36.8\%$. Without pre-training, cryoMAE shows reduced effectiveness, likely due to overfitting on the limited training data, hindering its generalization capabilities, especially in recognizing varied particle orientations and background noise variations.

\subsubsection{Max and mean matching strategies.}
Table~\ref{tab:matchingmethod} presents a comparative study on two similarity score calculation methods for matching sliding regions against exemplar latent features: maximum vs. average cosine similarity.
Table~\ref{tab:matchingmethod} reveals that maximum cosine similarity outperforms average cosine similarity, particularly in precision. This advantage is linked to the varied orientation distributions among particle exemplars. Maximum cosine similarity effectively matches regions to their closest exemplar across different orientations, ensuring optimal scores. Conversely, average cosine similarity dilutes scores for particles with diverse orientations, as it averages across all exemplars, including those with markedly different particle orientations from the target region. This dilution lowers similarity scores for such particles, reducing their distinctiveness from the background and making accurate particle identification more challenging amidst noise.

\subsection{Sensitivity Analysis}
In this section, we conducted a sensitivity analysis to examine the impact of varying the number of exemplars and the sliding stride on model performance.

\subsubsection{Number of exemplars.}

\begin{table}[!htb]
  \caption{CryoMAE with various exemplar number settings $\{1, 5, 15, 25\}$. }
  \label{tab:exemplar}
  \centering
  \resizebox{\textwidth}{!}{
  \begin{tabular}{@{}cccccccccccccc@{}}
    \toprule
     \multirow{2}{*}{EMPIAR ID}& \multicolumn{4}{c}{Precision} & \multicolumn{4}{c}{Recall} & \multicolumn{4}{c}{F1 Score} \\
    \cmidrule(l){2-5} \cmidrule(l){6-9} \cmidrule(l){10-13}
     & 1 & 5 & 15 & 25 & 1 & 5 & 15 & 25 & 1 & 5 & 15 & 25 & \\
    \midrule
    10081 & 0.167 & 0.401 & 0.645 & \textbf{0.653} & 0.774 & 0.896 & 0.939 & \textbf{0.943} & 0.275 &0.554&0.765 &\textbf{0.772}\\
    10093 & 0.068 & 0.243 & \textbf{0.383} & 0.373 & 0.139 & 0.361 & 0.497 & \textbf{0.508} & 0.091 &0.290&\textbf{0.433} &0.430\\
    10345 & 0.102 & 0.183 & \textbf{0.473} & 0.460 & 0.585 & 0.706 & 0.733 & \textbf{0.772} & 0.174 &0.291&0.575 &\textbf{0.576}\\
    10532 & 0.138 & 0.296 & \textbf{0.503} & 0.481 & 0.251 & 0.478 & 0.497 & \textbf{0.507} & 0.178 &0.366&\textbf{0.500} &0.494\\
    11056 & 0.287 & 0.289 & 0.694 & \textbf{0.712} & 0.353 & 0.485 & 0.671 & \textbf{0.680} & 0.317 &0.362&0.682 &\textbf{0.696}\\
    \midrule
    Average   & 0.152 & 0.282 & \textbf{0.540} & 0.536 & 0.420 & 0.585& 0.667 & \textbf{0.682} & 0.207 &0.373&0.591 &\textbf{0.594}\\
    \bottomrule
  \end{tabular}}
\end{table}

Table~\ref{tab:exemplar} shows how the model performance of cryoMAE varies with the number of exemplars used. As expected, adding more exemplars generally improves performance, owing to a more comprehensive representation of particle orientations in the similarity scoring process. This is particularly beneficial for particles with diverse orientations, as more exemplars increase the chance of capturing regions across different orientation states, improving recall. However, the performance improvement plateaus after a certain number of exemplars, with precision potentially decreasing. This is because particle orientations are limited, and once the diversity of these states is adequately covered, additional exemplars offer little benefit and may even raise false positives by increasing the likelihood of background regions being mistakenly scored highly. Thus, considering the diminishing returns beyond 15 exemplars, we identify this count as the optimal number for our few-shot learning approach.
\subsubsection{Sliding strides.}

\begin{table}[!htb]
  \caption{CryoMAE with various sliding strides \{14, 28, 42, 56\}. }
  \label{tab:stride}
  \centering
  \resizebox{\textwidth}{!}{
  \begin{tabular}{@{}cccccccccccccc@{}}
    \toprule
    \multirow{2}{*}{EMPIAR ID} & \multicolumn{4}{c}{Precision} & \multicolumn{4}{c}{Recall} & \multicolumn{4}{c}{F1 Score} \\
    \cmidrule(l){2-5} \cmidrule(l){6-9} \cmidrule(l){10-13}
     & 14 & 28 & 42 & 56 & 14 & 28 & 42 & 56 & 14 & 28 & 42 & 56 & \\
    \midrule
    10081 & 0.584 & 0.645 & 0.665 & \textbf{0.689} & \textbf{0.942} & 0.939 & 0.856 & 0.772 & 0.721 &\textbf{0.765}&0.748 &0.728\\
    10093 & 0.376 & 0.383 & 0.399 & \textbf{0.411} & 0.489 & \textbf{0.497} & 0.311 & 0.157 & 0.425 &\textbf{0.433}&0.350 &0.227\\
    10345 & 0.446 & 0.473 & 0.478 & \textbf{0.479} & \textbf{0.746} & 0.733 & 0.551 & 0.502 & 0.558 &\textbf{0.575}&0.512 &0.490\\
    10532 & 0.500 & 0.503 & 0.501 & \textbf{0.503} & \textbf{0.604} & 0.497 & 0.430 & 0.343 & \textbf{0.547} &0.500&0.463 &0.408\\
    11056 & 0.689 & 0.694 & 0.692 & \textbf{0.695} & \textbf{0.702} & 0.671 & 0.606 & 0.522 & \textbf{0.695} &0.682&0.646 &0.596\\
    \midrule
    Average   & 0.520 & 0.540 & 0.547 & \textbf{0.555} & \textbf{0.697} & 0.667& 0.551 & 0.459 & 0.589 &\textbf{0.591}&0.544 &0.490\\
    \midrule
    Average time (s)   & 356.4 & 87.2 & 38.0 & \textbf{20.8} & - & -& - & - & - &-&- &-\\
    \bottomrule
  \end{tabular}}
\end{table}

Table~\ref{tab:stride} outlines the model performance of cryoMAE across various sliding strides, noting that decreasing stride from 56 to 14 typically boosts recall but diminishes precision. This trend can be attributed to the fact that larger strides cause a certain particle to be present in fewer windows, minimizing duplicate detections and enhancing precision. However, this can result in lower similarity scores for many particles, as they're more likely to be close to window edges, which can reduce their likelihood of being selected and decrease recall. The F1 score, a precision-recall harmony measure, tends to improve with smaller strides. Yet, reducing stride size significantly lengthens processing time per query image. Considering the trade-off between time efficiency and model accuracy, a 28-pixel stride is identified as the optimal balanced approach.

\section{Conclusion}

We introduce cryoMAE, a pioneering approach in few-shot learning tailored specifically for the cryo-EM field, significantly reducing the dependence on extensive labeled datasets for accurate particle picking. By harnessing the power of MAE and integrating a novel self-cross similarity loss, cryoMAE achieves superior performance in identifying particle-containing regions amidst the challenges posed by low SNR and diverse particle orientations. Validations on the CryoPPP dataset demonstrate cryoMAE's superiority over existing NN-based methods, marking a significant advancement in the cryo-EM analysis pipeline. This innovation not only streamlines the process of high-resolution protein structure determination but also makes it more accessible to a wider scientific audience, promising to accelerate discoveries in structural biology.

\section{Acknowledgement}

This work was supported in part by U.S. NIH grants R01GM134020 and P41GM103712, NSF grants DBI-1949629, DBI-2238093, IIS-2007595, IIS-2211597, and MCB-2205148. This work was supported in part by Oracle Cloud credits and related resources provided by Oracle for Research, and the computational resources support from AMD HPC Fund. 

\newpage
\bibliographystyle{splncs04}
\bibliography{main}

\begin{thebibliography}{10}
\providecommand{\url}[1]{\texttt{#1}}
\providecommand{\urlprefix}{URL }
\providecommand{\doi}[1]{https://doi.org/#1}

\bibitem{bepler2019positive}
Bepler, T., Morin, A., Rapp, M., Brasch, J., Shapiro, L., Noble, A.J., Berger, B.: Positive-unlabeled convolutional neural networks for particle picking in cryo-electron micrographs. Nature methods  \textbf{16}(11),  1153--1160 (2019)

\bibitem{chen2020simple}
Chen, T., Kornblith, S., Norouzi, M., Hinton, G.: A simple framework for contrastive learning of visual representations. In: International conference on machine learning. pp. 1597--1607. PMLR (2020)

\bibitem{devlin2018bert}
Devlin, J., Chang, M.W., Lee, K., Toutanova, K.: Bert: Pre-training of deep bidirectional transformers for language understanding. arXiv preprint arXiv:1810.04805  (2018)

\bibitem{dhakal2023large}
Dhakal, A., Gyawali, R., Wang, L., Cheng, J.: A large expert-curated cryo-em image dataset for machine learning protein particle picking. Scientific Data  \textbf{10}(1), ~392 (2023)

\bibitem{gyawaliaccurate2023}
Gyawali, R., Dhakal, A., Wang, L., Cheng, J.: Accurate cryo-em protein particle picking by integrating the foundational ai image segmentation model and specialized u-net  (2023)

\bibitem{hadsell2006dimensionality}
Hadsell, R., Chopra, S., LeCun, Y.: Dimensionality reduction by learning an invariant mapping. In: 2006 IEEE computer society conference on computer vision and pattern recognition (CVPR'06). vol.~2, pp. 1735--1742. IEEE (2006)

\bibitem{he2022masked}
He, K., Chen, X., Xie, S., Li, Y., Doll{\'a}r, P., Girshick, R.: Masked autoencoders are scalable vision learners. In: Proceedings of the IEEE/CVF conference on computer vision and pattern recognition. pp. 16000--16009 (2022)

\bibitem{he2020momentum}
He, K., Fan, H., Wu, Y., Xie, S., Girshick, R.: Momentum contrast for unsupervised visual representation learning. In: Proceedings of the IEEE/CVF conference on computer vision and pattern recognition. pp. 9729--9738 (2020)

\bibitem{he2016deep}
He, K., Zhang, X., Ren, S., Sun, J.: Deep residual learning for image recognition. In: Proceedings of the IEEE conference on computer vision and pattern recognition. pp. 770--778 (2016)

\bibitem{iudin2023empiar}
Iudin, A., Korir, P.K., Somasundharam, S., Weyand, S., Cattavitello, C., Fonseca, N., Salih, O., Kleywegt, G.J., Patwardhan, A.: Empiar: the electron microscopy public image archive. Nucleic Acids Research  \textbf{51}(D1),  D1503--D1511 (2023)

\bibitem{li2022transfer}
Li, H., Chen, G., Gao, S., Li, J., Wan, X., Zhang, F.: A transfer learning-based classification model for particle pruning in cryo-electron microscopy. Journal of Computational Biology  \textbf{29}(10),  1117--1131 (2022)

\bibitem{milne2013cryo}
Milne, J.L., Borgnia, M.J., Bartesaghi, A., Tran, E.E., Earl, L.A., Schauder, D.M., Lengyel, J., Pierson, J., Patwardhan, A., Subramaniam, S.: Cryo-electron microscopy--a primer for the non-microscopist. The FEBS journal  \textbf{280}(1),  28--45 (2013)

\bibitem{pei2016simulating}
Pei, L., Xu, M., Frazier, Z., Alber, F.: Simulating cryo electron tomograms of crowded cell cytoplasm for assessment of automated particle picking. BMC bioinformatics  \textbf{17},  1--13 (2016)

\bibitem{punjani2017cryosparc}
Punjani, A., Rubinstein, J.L., Fleet, D.J., Brubaker, M.A.: cryosparc: algorithms for rapid unsupervised cryo-em structure determination. Nature methods  \textbf{14}(3),  290--296 (2017)

\bibitem{redmon2016you}
Redmon, J., Divvala, S., Girshick, R., Farhadi, A.: You only look once: Unified, real-time object detection. In: Proceedings of the IEEE conference on computer vision and pattern recognition. pp. 779--788 (2016)

\bibitem{rigort2012automated}
Rigort, A., G{\"u}nther, D., Hegerl, R., Baum, D., Weber, B., Prohaska, S., Medalia, O., Baumeister, W., Hege, H.C.: Automated segmentation of electron tomograms for a quantitative description of actin filament networks. Journal of structural biology  \textbf{177}(1),  135--144 (2012)

\bibitem{scheres2015semi}
Scheres, S.H.: Semi-automated selection of cryo-em particles in relion-1.3. Journal of structural biology  \textbf{189}(2),  114--122 (2015)

\bibitem{shi2022represent}
Shi, M., Lu, H., Feng, C., Liu, C., Cao, Z.: Represent, compare, and learn: A similarity-aware framework for class-agnostic counting. In: Proceedings of the IEEE/CVF Conference on Computer Vision and Pattern Recognition. pp. 9529--9538 (2022)

\bibitem{tang2007eman2}
Tang, G., Peng, L., Baldwin, P.R., Mann, D.S., Jiang, W., Rees, I., Ludtke, S.J.: Eman2: an extensible image processing suite for electron microscopy. Journal of structural biology  \textbf{157}(1),  38--46 (2007)

\bibitem{vinzenz2012actin}
Vinzenz, M., Nemethova, M., Schur, F., Mueller, J., Narita, A., Urban, E., Winkler, C., Schmeiser, C., Koestler, S.A., Rottner, K., et~al.: Actin branching in the initiation and maintenance of lamellipodia. Journal of cell science  \textbf{125}(11),  2775--2785 (2012)

\bibitem{voss2009dog}
Voss, N., Yoshioka, C., Radermacher, M., Potter, C., Carragher, B.: Dog picker and tiltpicker: software tools to facilitate particle selection in single particle electron microscopy. Journal of structural biology  \textbf{166}(2),  205--213 (2009)

\bibitem{wagner2019sphire}
Wagner, T., Merino, F., Stabrin, M., Moriya, T., Antoni, C., Apelbaum, A., Hagel, P., Sitsel, O., Raisch, T., Prumbaum, D., et~al.: Sphire-cryolo is a fast and accurate fully automated particle picker for cryo-em. Communications biology  \textbf{2}(1), ~218 (2019)

\bibitem{wang2016deeppicker}
Wang, F., Gong, H., Liu, G., Li, M., Yan, C., Xia, T., Li, X., Zeng, J.: Deeppicker: A deep learning approach for fully automated particle picking in cryo-em. Journal of structural biology  \textbf{195}(3),  325--336 (2016)

\bibitem{zeng2023high}
Zeng, X., Kahng, A., Xue, L., Mahamid, J., Chang, Y.W., Xu, M.: High-throughput cryo-et structural pattern mining by unsupervised deep iterative subtomogram clustering. Proceedings of the National Academy of Sciences  \textbf{120}(15),  e2213149120 (2023)

\bibitem{zhang2023genem}
Zhang, J., Chen, Q., Zeng, Y., Gao, W., He, X., Liu, Z., Yu, J.: Genem: Physics-informed generative cryo-electron microscopy. arXiv preprint arXiv:2312.02235  (2023)

\bibitem{zhu2017deep}
Zhu, Y., Ouyang, Q., Mao, Y.: A deep convolutional neural network approach to single-particle recognition in cryo-electron microscopy. BMC bioinformatics  \textbf{18},  1--10 (2017)

\end{thebibliography}

\end{document}